\documentclass[]{aa}
\usepackage{graphicx}
\begin{document}
\title{CCD photometric search for peculiar stars in open clusters. VI. 
NGC~1502, NGC~3105, Stock~16, NGC~6268, NGC~7235 and NGC~7510\thanks{Based on 
observations at the Asiago observatory, BNAO Rozhen, CTIO (Proposal
2003A-0057), ESO-La Silla (Proposal 073.C-0144)}}
\author{E.~Paunzen\inst{1}, M.~Netopil\inst{1}, I.Kh.~Iliev\inst{2}, H.M.~Maitzen\inst{1}, 
A.~Claret\inst{3}, O.I.~Pintado\inst{4}}

\mail{Ernst.Paunzen@univie.ac.at}

\institute{Institut f{\"u}r Astronomie der Universit{\"a}t Wien,
           T{\"u}rkenschanzstr. 17, A-1180 Wien, Austria
\and	   Institute of Astronomy, National Astronomical Observatory, 
           P.O. Box 136, BG-4700 Smolyan, Bulgaria
\and	   Instituto de Astrof{\'i}sica de Andaluc{\'i}a
		   CSIC, Apartado 3004, 18080 Granada, Spain
\and	   Departamento de F{\'i}sica, Facultad de Ciencias Exactas 
           y Tecnolog{\'i}a, Universidad Nacional de Tucum{\'a}n, Argentina - Consejo Nacional 
		   de Investigaciones Cient{\'i}ficas y T{\'e}cnicas de la Rep{\'u}blica Argentina}

\date{Received 22 April 2005 / Accepted 21 July 2005}
\authorrunning{E. Paunzen et al.}{}
\titlerunning{Photometric search for peculiar stars in open clusters. VI.}{}

\abstract{In a sample of six young open clusters (NGC~1502, NGC~3105, Stock~16, 
NGC~6268, NGC~7235, and NGC~7510) we investigated 1753 objects using the narrow band,
three filter 
$\Delta a$ photometric system resulting in the detection of eleven bona-fide magnetic
chemically peculiar (CP) stars 
and five Be or metal-weak stars. The results for the distant cluster NGC~3105 is most 
important because of the still unknown influence of the global metallicity gradient of the 
Milky Way. These findings confirm that CP stars are present in open clusters of very young 
ages (log\,$t$\,$\geq$\,6.90) at galactocentric distances up to 11.4\,kpc. For all programme 
clusters the age, reddening, and distance modulus were derived using the corresponding isochrones. 
Some additional variable stars within Stock~16 could be identified by comparing 
different photometric studies.

\keywords{Stars: chemically peculiar -- stars: early-type -- techniques:
photometric -- open clusters and associations: general}
}
\maketitle

\section{Introduction}

We have observed very young objects,
including pre-main-sequence (PMS) stars,
and distant open clusters in order to detect classical chemically peculiar
(CP) stars of the upper main sequence. We applied the 
$\Delta a$ photometric system
that measures the flux depression at 5200\,\AA, a typical 
feature of CP and related objects (Kupka et al. 2004).

The detection of CP stars in young open clusters that are at significantly 
different galactocentric distances to the Sun will help us understand
the evolution and formation of such objects in non-solar environments. 

The data were collected at four different observatories and thus with widely
different instruments and CCD detectors. Because this and the 
use of different Johnson $V$ sources, ranging from photographic to CCD observations, 
the absolute transformation coefficients from our observed $y$ to Johnson $V$ 
magnitudes vary significantly, but the final color-magnitude-diagrams of 
all programme clusters are fully compatible. The same holds for the 
derived normality lines. This is further proof of the intrinsic
consistency of the CCD $\Delta a$ photometric system.

In addition, we applied the isochrones for the $\Delta a$ photometric
system (Claret et al. 2003) that allows us to determine the age, reddening and
distance modulus with an appropriate accuracy. The results from our isochrone
fitting procedure were compared with published parameters yielding an excellent
agreement. 

We detected five CP stars in NGC~1502, Stock~16 and NGC~7235 that have
ages less than 10\,Myr as well as five objects with significant
positive $\Delta a$ values in NGC~6268 (40\,Myr). 
The results for NGC~3105, where one CP star was found, is most important because its
distance from the Sun is about 8.5\,kpc with a galactocentric distance of 11.4\,kpc.

\section{Observations, reduction and methods}

Observations of the six open clusters were performed 
at four different sites and telescopes: 
\begin{itemize}
\item 1.82\,m telescope (Cima Ekar, Asiago), AFOSC, TK1024AB 1024\,$\times$\,1024 pixel
CCD, 8' field-of-view
\item 2\,m RCC telescope (BNAO, Rozhen),
direct imaging, SITe SI003AB 1024\,$\times$\,1024 pixel CCD,
5' field-of-view
\item 0.9\,m telescope (CTIO), direct imaging, SITe 2084\,$\times$\,2046 pixel CCD,  
13' field-of-view
\item 3.6\,m telescope (ESO-La Silla), EFOSC2, Loral/Lesser 2048\,$\times$\,2048 pixel
CCD, 5' field-of-view
\end{itemize}
The observing log with the number of frames in each filter
is listed in Table \ref{log}. The observations were performed with
two different filter sets, both having the following characteristics:
$g_1$ ($\lambda_c$\,=\,5007\,\AA, FWHM\,=\,126\,\AA, $T_P$\,=\,78\%), 
$g_2$ (5199, 95, 68) and $y$ (5466, 108, 70).

The basic CCD reductions and point-spread-function-fitting 
were carried out with standard IRAF V2.12.2 routines on
Personal Computers running under LINUX.
For some clusters we also checked these procedures by
applying aperture photometry with excellent
agreement.
The method of calculating the normality line, deriving the errors
as well as the calibration of our $(g_1-y)$ as well as $y$ measurements, 
is the same as in previous works (Bayer et al. 2000; Paunzen et al. 2003
and references therein) and will not repeated here.  
 
\begin{table}[t]
\begin{center}
\caption{Observing log for the programme clusters. All clusters were observed on one night by 
I.Kh.~Iliev (II), H.M.~Maitzen (HM), and M.~Netopil (MN).}
\label{log}
\begin{tabular}{lcccccc}
\hline\hline
Cluster & Site & Date & Obs. & \#$_{g_{\rm 1}}$ & \#$_{g_{\rm 2}}$ & \#$_{y}$ \\
\hline
NGC~1502 & Asiago & 01.2004 & MN & 8 & 8 & 12 \\
NGC~3105 & CTIO & 04.2003 & HM & 6 & 6 & 6 \\
Stock~16 & ESO & 06.2004 & MN & 10 & 10 & 11 \\
NGC~6268 & ESO & 06.2004 & MN & 10 & 10 & 11 \\
NGC~7235 & BNAO & 09.2004 & II & 10 & 10 & 10 \\
NGC~7510 & BNAO & 09.2004 & II & 12 & 12 & 12 \\
\hline
\end{tabular}
\end{center}
\end{table}

The isochrones shown in Fig. \ref{cmds} are based on the
$\Delta a$ photometric system and were taken from
Claret et al. (2003). The derived ages, reddening and
distance moduli together with the errors are listed in Table \ref{all_res}.
The fitting procedure takes advantage of the available $UBV$ measurements for
all programme clusters by comparing our results to those 
of the color-magnitude-diagrams for the $UBV$ photometric system.
However, our determination is based on the $\Delta a$ measurements
only, which is another important application of this photometric system.

\begin{table*}[t]
\begin{center}
\caption{Summary of results; the age, distance modulus, reddening
and thus the distance from the Sun was derived by fitting isochrones
to the $\Delta a$ photometry. $R_V$ was set to 3.1, except for NGC~1502,
for which Pandey et al. (2003) list 2.57. The distance of the Sun from
the galactic center $R_0$ was set to 8.5\,kpc.
The errors in the final digits of the corresponding quantity
are given in parenthesis.}
{\scriptsize
\label{all_res}
\begin{tabular}{lcccccc}
\hline\hline
Name & NGC 1502 & NGC 3105 & Stock 16 & NGC 6268 & NGC 7235 & NGC 7510 \\
     & C0403+622 & C0959$-$545 & C1315$-$623 & C1658$-$396 & C2210+570 
	 & C2309+603 \\
\hline
$l/b$ & 143.7/+7.7 & 279.9/+0.3 & 306.1/+0.1 & 346.1/+1.2 & 102.7/+0.8 
      & 111.0/+0.0 \\
$E(B-V)$ ($\pm$0.02) & 0.75 & 0.95 & 0.52 & 0.40 & 0.90 & 0.90 \\
$m_V - M_V$ ($\pm$0.2) & 12.1 & 17.6 & 12.9 & 11.4 & 15.4 & 15.5 \\
$d$\,[pc] & 1080(130) & 8530(1030) & 1810(220) & 1080(130) & 3330(400) & 3480(420) \\
$R_{GC}$\,[kpc] & 9.4(1) & 11.4(6) & 7.6(1) & 7.5(2) & 9.8(2) & 10.3(2) \\
$|z|$\,[pc] & 145(17) & 45(5) & 3(1) & 23(2) & 46(6) & 0(1) \\
log\,$t$ ($\pm$0.1) & 6.90 & 7.30 & 6.90 & 7.60 & 6.90 & 7.35 \\
Tr-type & I 3 m & I 3 p & IV 2 p & II 2 p & III 2 p & II 2 m \\
n(mem) & 34 & 48 & 23 & 34 & 87 & 100 \\
n(none) & 6 & 1257 & 70 & 28 & 42 & 24 \\
CP (No./WEBDA) & 1/27 & 617/$-$ & 15/12 & 15/80  & 2/$-$   & 48/65 \\
               &      &         &       & 40/39  & 96/$-$  & 121/27 \\
			   &      &         &       & 63/21  & 110/$-$ & \\
			   &      &         &       & 70/23  & 121/18  & \\
			   &      &         &       & 82/$-$ & 123/90  & \\
			   &      &         &       &        & 127/92  & \\
$\Delta$a/$(B-V)_0$/$M_V$ & +88/$-$240/+370 & +31/+40/$-$740 & +24/$-$138/+480 
& +56/$-$62/+550  & +56/+119/+1520 & $-$46/$-$18/$-$1020 \\
\,[mmag] &          &      &              & +23/$-$116/+760 & +48/+176/+1680 & $-$27/+209/$-$3690 \\
         &          &      &              & +19/$-$112/$-$220 & +31/$-$86/+1070 \\
         &          &      &              & +12/$-$112/+2130 & $-$41/$-$53/$-$1730 \\
         &          &      &              & +18/+21/+1500    & $-$54/+207/$-$210 \\
         &          &      &              &                  & $-$39/+116/+2160 \\					
n(frames) & 28 & 18 & 31 & 31 & 30 & 36 \\
\hline
\end{tabular}
}
\end{center}
\end{table*}

The tables with all data for the individual cluster stars as well as
nonmembers are available
in electronic form at the CDS via anonymous ftp to cdsarc.u-strasbg.fr (130.79.125.5),
http://cdsweb.u-strasbg.fr/Abstract.html
or upon request from the first author. These tables include the cross
identification of objects from the literature, the $X$ and $Y$
coordinates of our frames, the observed $(g_{1}-y)$ and
$a$ values with their corresponding errors, $V$ magnitudes,
the $(B-V)$ colors from the literature,
$\Delta a$-values derived from the normality lines of $(g_{1}-y)$,
(exclusive nonmembers) and the number of observations, respectively.

\begin{figure*}
\begin{center}
\includegraphics[width=150mm]{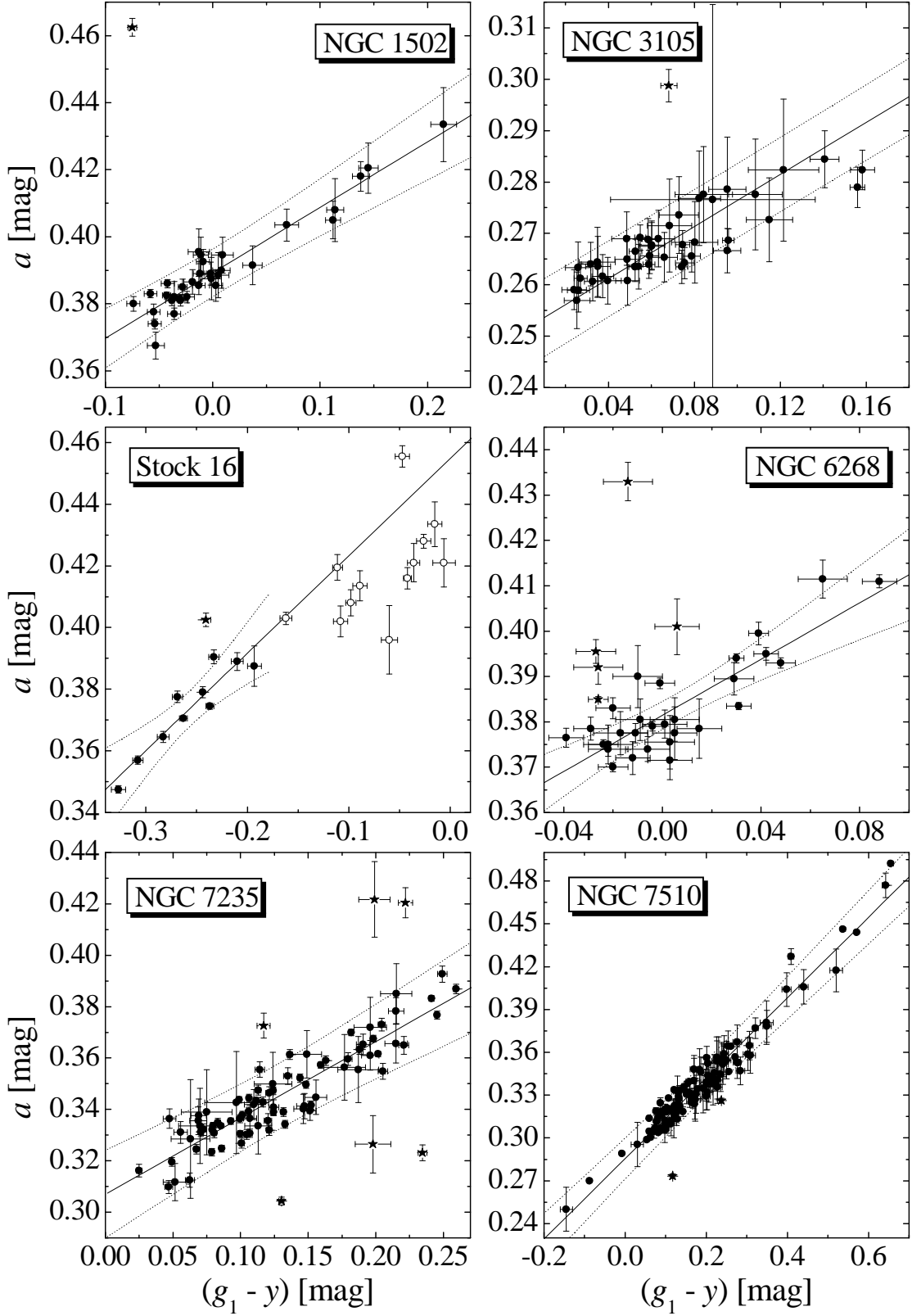}
\caption[]{Observed $a$ versus $(g1-y)$
diagrams for our programme clusters. The solid line is the
normality line and the dotted lines are the confidence intervals
corresponding to 99.9\,\%. The error bars for each individual object
are the mean errors. The detected peculiar objects are marked with
asterisks. Only members (filled circles) have been used to derive the
normality lines. For Stock~16, we also included the PMS objects
(open circles)
to show the apparent emission (location below the normality line) for
these stars. The fitting parameters are listed in Table \ref{coeffs}.}
\label{normalities}
\end{center}
\end{figure*}

\begin{figure*}
\begin{center}
\includegraphics[width=155mm]{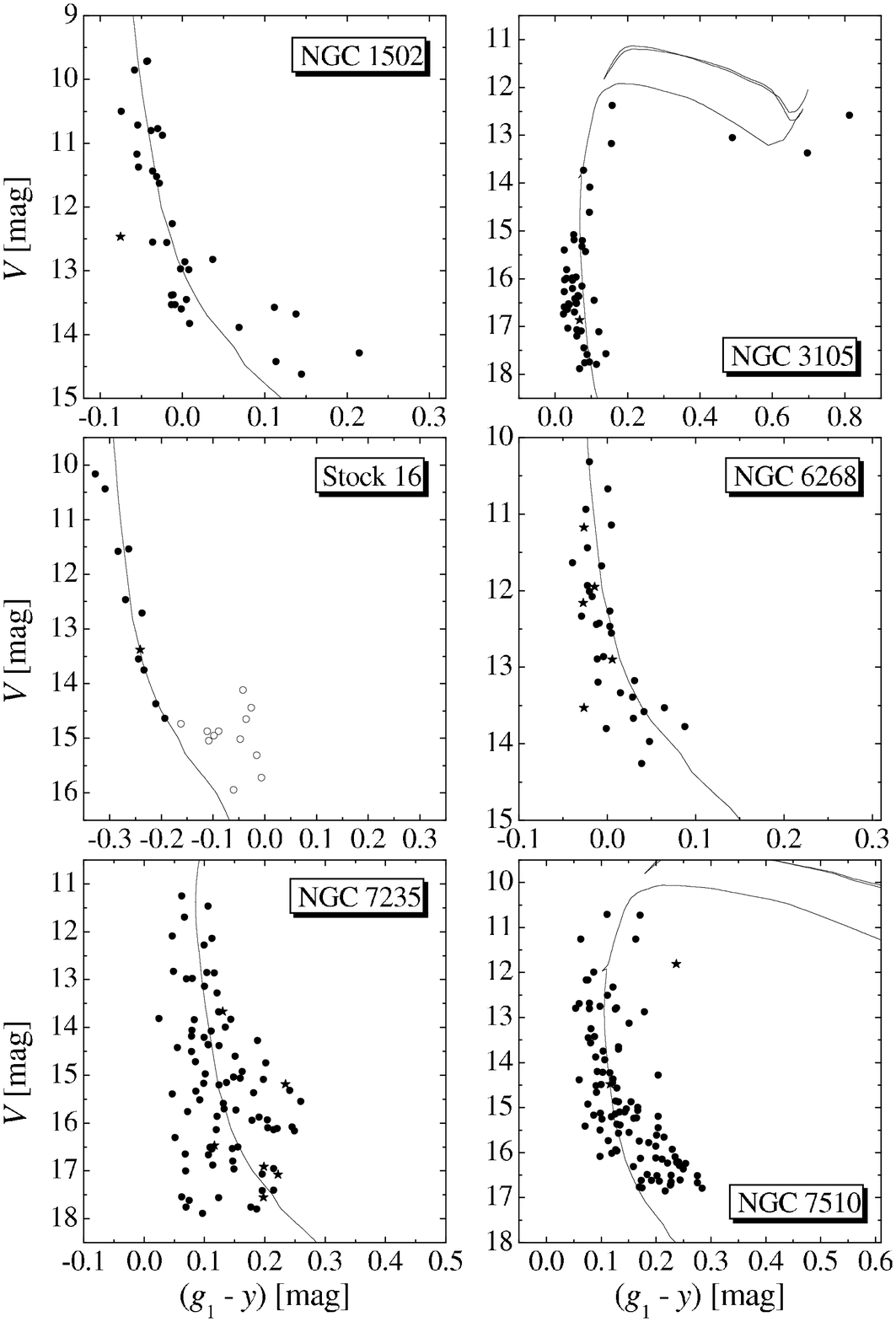}
\caption[]{Observed $V$ versus $(g1-y)$
diagrams for our programme clusters. The isochrones 
are based on the $\Delta a$ photometric system and were
taken from Claret et al. (2003). The derived ages, reddening, and
distance moduli are listed in Table \ref{all_res}. The symbols
are the same as in Fig. \ref{normalities}.}
\label{cmds}
\end{center}
\end{figure*}

The diagnostic diagrams for
all six open clusters are shown in Fig. \ref{normalities}. Also,
the normality lines and the confidence intervals corresponding to 99.9\,\%
are plotted. The detected peculiar objects are marked with
asterisks. Only members (filled circles) have been used to derive the
normality lines. The selection of these objects are according to their
location in the color-magnitude-diagrams as well as the distance from the
cluster centers and additional information
from the literature (proper motions and radial velocities) taken from
WEBDA (http://obswww.unige.ch/webda/).

\section{Results} \label{results}

In the following we will discuss the results and the comparison with
the literature for the individual open clusters in more detail.

{\it NGC~1502:} There are several photometric studies of this cluster
in the literature (Purgathofer 1961; Reimann \& Pfau 1987; Tapia et al. 1991;
Delgado et al. 1992; Crawford 1994), but except 
Purgathofer (1961), all studies are limited to about 12th magnitude. A study of
proper motions and the membership probability was carried out by 
Hopmann (1958).
Our age determination for NGC~1502 of log\,$t$\,=\,6.9 agrees very well with 
the values found in the literature ranging from 6.7 (Tapia et al.
1991) to 7.0 (Reiman \& Pfau 1987). It is therefore a very young open
cluster within the galactic disk.  
According to Tapia et al. (1991) and Pandey et al. (2003) 
the reddening law towards NGC~1502 is anomalous with values of $R_V$ between
2.42 ans 2.57. Using a value of 2.57 and an $E(B-V)$\,=\,0.75\,mag,
we obtain a true distance modulus $(m_V-M_V)_0$\,=\,10.17 and 
therefore a distance of 1080\,pc from the Sun.
The apparent distance modulus of 12.1 is in line with the values published
by Crawford (1994, 12.02) and Purgathofer (1961, 12.0). Tapia et al. (1991), on the
other hand, found a value of 12.7. They report some inconsistencies when calculating
the mean absorption towards NGC~1502. This might be the reason for the deviating
value.
One CP2 candidate was found (\#27) with $\Delta a$\,=\,+88\,mmag which is one of the most
extreme values observed yet. This object was also recognized as
peculiar within the Geneva photometric system with $\Delta(V1-G)$\,=\,+58\,mmag 
(North \& Cramer 1981). This star shows an extreme ``blueing'' effect which
is typical for some magnetic CP objects  
due to stronger UV absorption than in normal type stars (Adelman 1980). 
However, its 
cluster membership is confirmed by the proper motion and the analysis
by Tapia et al. (1991) who used $JHK$ and $uvby\beta$ 
photometry. 
\\
{\it NGC~3105:} we have included this open cluster because of the discrepant distances 
from 5.5(8) to 9.5(1.5)\,kpc found in the literature (Sagar et
al. 2001) and its young age. This implies that NGC~3105 has only a small apparent
diameter on the sky (1.5') and therefore a large number of measured nonmembers. 
Sagar et al. (2001) list a reddening $E(B-V)$\,=\,1.06\,mag and an age
of log\,$t$\,=\,7.40(25). We find a slightly lower mean reddening (0.95) but
a comparable age (7.30). Figure \ref{cmds} shows that the coolest red giants
cannot be fitted by isochrones in terms of the color 
also clearly visible in Sagar et al. (2001, Fig. 19 therein). 
These objects likely show an extended atmosphere with strong 
stellar winds and mass-loss which severely influences the observed colors
(Eigenbrod et al. 2004). However this effect does not influence the fitting
of the isochrone itself. There is no doubt that NGC~3105 is a very distant
open cluster ($d$\,=\,8.53(1.03)\,kpc) that includes at least one CP star 
(\#617, $\Delta a$\,=\,+31\,mmag). Figure \ref{cmds} shows that this object
is a definite member with an apparent distance of 0.68' from the clusters
center. Unfortunately, it was not measured by Sagar et al. (2001).
The finding of one CP star in an open cluster with a distance of 11.4(6)\,kpc
from the galactic center is most important because of the still unknown
influence of the global metallicity gradient of the Milky Way (Chen et al. 2003) on the 
formation and evolution of CP stars.
\\
{\it Stock~16:} Vazquez et al. (2005) recently identified 27 members
on the basis of a deep CCD $UBV(RI)_c$ photometric study. This open 
cluster is very young with an age of log\,$t$\,=\,6.9 derived from our
isochrone fitting which is in line with the results from Turner (1985,
6.5--6.7) and Vazquez et al. (2005, 6.7--6.8). There is a significant
number of PMS objects identified in the
literature (Fig. \ref{cmds}). Most of these PMS stars lie below the 
normality line (Fig. \ref{normalities}) 
most certainly caused by the emission of these objects which is   
a well known phenomenon (Reipurth et al. 1996). The normality 
line was therefore calculated using only the definite members excluding 
the possible PMS stars. Vazquez et al. (2005) mentioned photometric 
discrepancies of some stars compared to the paper by Turner (1985).
We found several other stars showing deviations of more 
than 0.2\,mag between the different photometric studies which might be 
caused by the intrinsic variability of PMS objects on various 
time scales with amplitudes of the same level as the detected
deviations (Zwintz et al. 2005). We compared the available
photometry from the literature and searched for objects with at
least three significant deviating observations. 
This resulted in the unambiguous variability for the objects No. 1, 10 
and 100 (numbering system according to WEBDA). Misidentification
in the literature can be excluded because all stars are well separated
in the field of Stock~16.
A detailed time series analysis of our data will be presented in a separate
paper. We discovered one CP candidate (WEBDA \#12, $\Delta a$\,=\,+24\,mmag)  
in the investigated sample of stars. However, the membership of this object
is controversial. Vazquez et al. (2005) defined this star as a probable nonmember, 
whereas Fenkart et al. (1977) and Turner (1985) classified it as a definite member.
From the location of it in various $RGU$ and $UBV(RI)_c$ photometric diagrams, 
we conclude that this object seems to be a member.  
\\
{\it NGC~6268:} this is the most poorly investigated cluster in our sample. 
The only published study, by Seggewiss (1968), is based on photographic plates and
shows an excellent agreement for the distance modulus (11.46 versus
11.40) and reddening (0.41 and 0.40) compared to the isochrone fitting procedure presented 
in this analysis. He stated that there are no evolved members, i.e. giants in this
open cluster. We have compared our $(g_1-y)$ data with the photographic
$(B-V)$ and found a large scatter.
Lyng\aa\,(1987) refers to Moffat \& Vogt (1975) and a published age of log\,$t$\,=\,7.4 
(we have obtained 7.6), but this value could not be retraced, since in the original paper 
no explicit parameters for NGC~6268 have been given.
Because of the very low $\Delta a$ detection limit, several apparent 
CP stars were identified. At least three objects show a $\Delta a$ value of more than +20\,mmag
with an extreme value of +59\,mmag. This situation is similar to the results
found for NGC~2516 (log\,$t$\,=\,7.4, P{\"o}hnl et al. 2003) which hosts a significant 
number of CP stars
with a large range of peculiarity degrees (Maitzen \& Hensberge 1981). Bagnulo et al. (2003)
detected a 14.5\,kG magnetic field for HD~66318 which is a member of NGC~2516.
So further spectroscopic investigations of the bona-fide CP candidates for NGC~6268 are
needed. 
\\
{\it NGC~7235:} Pigulski et al. (1997) presented an extensive study of this
very young (log\,$t$\,=\,6.9)
open cluster on the basis of $BV(RI)_c$ photometry and additional H$\alpha$
measurements. They included time series for nine variable stars of all
kinds. We are able to confirm the variable stars in common a detailed
analysis of those objects will be published elsewhere.
Chopinet (1956) classified WEBDA \#1 as A1\,Ia\,p, but other
sources list B9\,Iab (Hiltner 1956), B8\,Ia (Sowell 1987) and B8\,I 
(Massey et al. 1995). Because this object has a $V$ magnitude of 8.8, we 
were not able to measure a $\Delta a$ index. However, this object is
probably not a classical CP star but an evolved supergiant. We found six
objects with significant deviating $\Delta a$ values
(three with positive and three with negative). One star (WEBDA \#18)
seems to be a Be object whereas the other ones (\#90 and \#92) are good
candidates for metal-weak objects because they are too cool to be B-type
stars. The three objects with significant $\Delta a$-values are most
certainly CP stars. All objects seem to be members of NGC~7235.
\\
{\it NGC~7510:} no classical CP star was detected in this cluster, but we
are able to find significant negative $\Delta a$ values for the previously known 
(Sagar \& Griffiths 1991, Barbon \& Hassan 1996) Be objects WEBDA \#27 and \#65 which 
is typical of their emission phase (Fig. \ref{normalities}).
Again, this is indication that the $\Delta a$ photometric system is able to
detect Be stars with high efficiency. Our derived age, reddening and distance modulus is 
comparable to the literature values placing it in the Perseus arm of the Milky Way.

\begin{table}[t]
\begin{center}
\caption{The regression coefficients for the
transformations and normality lines. The absolute values and errors
vary due to the inhomogeneous ``standard'' observations (photographic, photoelectric,
and CCD) found in the literature as well as the dependence on the 
magnitude range in common, i.e. a broader range guarantees a small error. 
The offsets are due to the four different telescopes and thus instruments as well
as CCD used (Table \ref{log}).
The errors in the final digits of the corresponding quantity
are given in parenthesis.}
\label{coeffs}
\begin{tabular}{ll|l}
\hline\hline
Cluster & $V$\,=\,a\,+\,b$\cdot(y)$, $N$ & $a_{0}$\,=\,a\,+\,b$\cdot(g_{1}-y)$, $N$ \\
\hline 
N~1502 & +0.70(10)/0.978(9)/21 & 0.389(1)/0.195(11)/33 \\
N~3105 & $-$5.52(24)/1.01(1)/115 & 0.251(1)/0.253(4)/47 \\
Stock~16 & $-$0.07(9)/1.023(8)/8 & 0.46(1)/0.316(43)/10 \\
N~6268 & +0.53(9)/0.937(8)/31 & 0.381(1)/0.310(37)/29 \\
N~7235 & $-$1.81(22)/0.89(1)/67 & 0.307(2)/0.297(15)/81 \\
N~7510 & $-$0.07(15)/0.78(1)/122 & 0.286(1)/0.282(6)/122 \\
\hline
\end{tabular}
\end{center}
\end{table}

\section{Conclusions}

We detected eleven bona-fide chemically peculiar stars, five (two previously identified)
Be stars as well as metal-weak stars in six young open clusters of the Milky Way. These 
results are based on photometric $\Delta a$ measurements of 174 individual frames from 
four different observatories.

As an important application of the $\Delta a$ photometric system, 
isochrones were fitted to the color-magnitude-diagrams ($V$ versus
$(g_1-y)$) of the programme clusters. For this purpose, our measured
$y$ magnitudes were directly converted into standard $V$ magnitudes
on the basis of already published values. A comparison of our results
yields an excellent agreement with the appropriate parameters from
the literature. 

These findings confirm that CP stars are present in open clusters of
very young ages (log\,$t$\,$\geq$\,6.90) at galactocentric distances up
11.4\,kpc.

\begin{acknowledgements}
We thank our referee, Dr. Glagolevskij for useful comments.
This research was performed within the projects  
{\sl P17580} and {\sl P17920} of the Austrian Fonds zur F{\"o}rderung der 
wissen\-schaft\-lichen
Forschung (FwF) and benefited also from the financial contributions of the City of
Vienna (Hochschuljubil{\"a}umsstiftung project: $\Delta a$ Photometrie in der 
Milchstrasse und den Magellanschen Wolken, H-1123/2002). The observations at ESO were 
funded by the Optical Infrared Coordination network (OPTICON),
a major international collaboration supported by the
Research Infrastructures Programme of the
European Commission's Sixth Framework Programme.
One of us (I.Kh. Iliev) acknowledges the partial support by the Bulgarian National Science Fund
under grant F-1403/2004.
Use was made of the SIMBAD database, operated at the CDS, Strasbourg, France and
the WEBDA database, operated at the Institute of Astronomy of the University
of Lausanne. This research has made use of NASA's Astrophysics Data System.
\end{acknowledgements}

\end{document}